\documentclass[preprint,proceedings]{rmaa}
\usepackage{paralist}

\usepackage{psfrag,color}

\def\mpch{\mbox{h$^{-1}$Mpc}}
\def\rvir{\mbox{R$_{vir}$}}
\def\kms{\mbox{kms$^{-1}$}}

\def\sig{\mbox{$\sigma_{DM}$}}
\def\rhosl{\mbox{$\rho_{c,-1}$}}
\def\sign{\mbox{$\sigma_0$}}
\def\vmax{\mbox{v$_{\rm max}$}}
\def\vrel{\mbox{v$_{12}$}}
\def\vcien{\mbox{v$_{100}$}}
\def\ncol{\mbox{N$_{\rm col}$}}
\def\densunit{\mbox{h$^2$M$_{\odot}$Mpc$^{-3}$}}
\def\LCDM{\mbox{$\Lambda$CDM}}

\SetYear{2002}
\SetConfTitle{Galaxy Evolution: Theory and Observations}

\RescaleTitleLengths{0.7}
\title{The structure of halos in Self-Interacting Cold Dark Matter models} 

\author{P. Col\'{\i}n,\altaffilmark{1}
        V. Avila-Reese,\altaffilmark{2}
        O. Valenzuela,\altaffilmark{3}
        C. Firmani\altaffilmark{4}} 

\altaffiltext{1}{Instituto de Astronom\'\i{}a, UNAM, Morelia, M\'exico.}
\altaffiltext{2}{Instituto de Astronom\'\i{}a, UNAM, CU, M\'exico.}
\altaffiltext{3}{Astronomy Department, New Mexico State University}
\altaffiltext{4}{Osservatorio Astronomico di Brera, Merate, Italy}

\shortauthor{Col\'{\i}n et al.}
\shorttitle{Self-Interacting Dark Matter }

\fulladdresses{
\item P. Col\'{\i}n: 
  Instituto de Astronom\'\i{}a, Universidad Nacional
  Aut\'onoma de M\'exico, Campus Morelia, Apartado Postal 3--72, 58090
  Morelia, Michoac\'an, M\'exico (p.colin@astrosmo.unam.mx).
}

\listofauthors{P. Col\'{\i}n, V. Avila-Reese, O. Valenzuela, \& C. Firmani }
\indexauthor[Colin, P.]{Col\'{\i}n, P.}
\indexauthor{Avila-Reese, V.}
\indexauthor{Valenzuela, O.}
\indexauthor{Firmani, C.}
\abstract{High-resolution numerical simulations were performed
to study the structure and substructure of Milky Way- and cluster-sized 
halos in a \LCDM\ cosmology with self-interacting 
(SI) dark particles, where the particle cross section, \sig, is assumed
constant or inversely proportional to the relative velocity. 
We conclude that the cuspy halo problem at galaxy scales of the \LCDM\ 
cosmogony can be solved in the latter case. At the same time, the subhalo 
population remains roughly similar to that seen on CDM halos.
}

\resumen{Usando simulaciones de alta resoluci\'on, hemos estudiado la estructura 
y subestructura de halos correspondientes a galaxias tipo V\'{\i}a L\'actea y 
a c\'umulos de galaxias para una cosmolog\'{\i}a \LCDM\ con part\'{\i}culas 
oscuras auto-interactuantes. La secci\'on eficaz, \sig, en nuestros
modelos puede ser constante o proporcional al inverso de la velocidad
relativa, v$_{12}$. Esta \'ultima dependencia de \sig\ con v$_{12}$
resuelve el problema del ``n\'ucleo empinado'' de la
cosmogon\'{\i}a \LCDM\ y predice una poblaci\'on de subhalos similar a 
la que se observa en los halos CDM.}

\addkeyword{cosmology:dark matter}
\addkeyword{galaxies:halos}
\begin{document}
% Typeset article header
\maketitle

\section{Introduction}
\label{sec:intro}

The potential problems of the popular \LCDM\ model at small
scales have motivated the search for modifications that keep
the succesful predictions of the model at large scales 
unaltered. 
Two of these problems are that (a) \LCDM\ halos seem to be more 
concentrated and cuspy than what is suggested by the rotation curves 
of dwarf and LSB galaxies (see references below), and that
(b) the number of subhalos predicted by the \LCDM\ model on
galactic scales overwhelms the observed one (Klypin et al. 1999). 
While the latter problem has a well-motivated 
astrophysical solution (see e.g., Benson et al. in
this volume), the former one seems to require a solution
which goes beyond the standard model of structure
formation. A posible solution for the ``cuspy'' problem
was proposed by Spergel \& Steinhard (2000) (see also
Firmani et al. 2000) who introduced the concept of
self-interacting (SI) DM particles within this context.
The SIDM cross section per unit of particle mass, \sig, is
a free parameter which may be constant or dependent on the 
relative velocity of the colliding particles, 
\vrel: $\sig = \sign \vcien^{-\alpha}$, where \vcien\ is \vrel\
in units of 100 \kms.

A number of authors have constrained the range of 
the pair ($\sign,\alpha$) of values and concluded that the relevant
regime for structure formation is the optically thin one (small 
cross sections; see e.g., Hennawi \& Ostriker 2002 and referenes therein).
By means of cosmological simulations, we explore
in more detail the effects of varying ($\sign,\alpha$) on halo 
structure and subhalo population (Col\'{\i}n et al. 2002, hereafter C2002 ). 
An outline of the results is presented in \S 3. In order to explore the viability of the 
SIDM model, it is important to establish the observational 
constraints. This is done in the next section.

\section{Observational constraints}

Several recent observational studies, using the highest sensitivity 
and spatial 
resolution in HI, H$\alpha$ and CO lines, show that halos of dwarf and LSB 
galaxies seem to be less concentrated in the center than what is predicted 
by the \LCDM\ model (e.g., C{\^ o}t{\' e}, Carignan, \& Freeman 2000; 
Blais-Ouellette, Amram, \& Carignan 2001; 
Bolatto et al. 2002; Amram \& Garrido 2002; de Blok et al. 2001a, 2001b;
Marchesini et al. 2002;
see also the contributions by Bosma and by de Blok in this volume). 
The inner slopes inferred for the halo density distribution are typically
around $-0.5$, shallower than the slopes of \LCDM\ halos. 
There are also pieces of evidence of soft cores in normal disk galaxies 
(e.g., Corsini et al. 1999; Firmani \& Avila-Reese 2000; Borrielo \& Salucci 2001; 
Salucci 2001) and in elliptical galaxies (Keeton 2001), although 
the evidence is less direct than in the case of dwarf and LSB galaxies.

An important question is that of the scaling laws of halo cores.
Assuming the non-singular isothermal or pseudo-isothermal halo models,
several authors have tried to find this law.  For a sample of dwarf and 
LSB galaxies with high-quality rotation curves and two clusters of galaxies 
studied with gravitational lensing, Firmani et al. (2000, 2001) have found that 
the halo central density, when plotted against the maximum circular velocity, 
\vmax, exhibits a large scatter with no correlation with \vmax; the core radius,
on the other hand, tend to correlate with \vmax\ with a slope smaller than 1.
Other workers have arrived 
to similar conclusions (see references in C2002). The data on
clusters have increased in the last two years, mainly from high-resolution Chandra 
X-ray studies. The mass distributions for most of the X-ray clusters
are well fitted by both the NFW and the pseudo-isothermal profiles.
The studies with strong gravitational lensing favour shallow halos.

For a sample of dwarf and LSB galaxies with high-resolution rotation
curves and for X-ray and lensing clusters of galaxies, we infer that 
the central density \rhosl, defined as the density 
where the logarithmic slope of the density profile 
becomes lower than $-1$, has values
between  $\sim 10^{16}-10^{17}\ \densunit$ from dwarf- to
cluster-sized halos, with no evidence of a dependence on \vmax. The 
predictions of the \LCDM\ model for \rhosl\ for cluster-sized halos
are within this range, but for galaxy-sized halos, \rhosl\ is larger
than $10^{17}\ \densunit$. Can the SIDM cosmology solve the discrepancy?

\section{Models and results}

The Adaptive Refinement Tree (ART) N-body code (Kravtsov, Klypin, \& Khokhlov 1997) 
has been used to run the N-body simulations. A pair of particles
collide if the distance $d=- \lambda \ln(1 - P)$ ($\lambda = 1 
/ \rho \sig$ is the mean free path and $P$ is a random number distributed 
uniformly between 0 and 1) becomes lower than the distance
$\vrel \Delta t$, where \vrel\ is the relative velocity between the
particle and one of its nearest neighbors, and $\Delta t$ is the 
time step. The \LCDM\ model was used throghout.

%physics
The core evolutionary behavior of NFW halos with the self-interaction on is shown 
in our experiments with monolithic halos. Firstly, the core expands due 
to the heat inflow from the hotter surroundings (for CDM halos, the 3D 
velocity dispersion decreases towards the center). There is a radius
below which quantities like the total energy or the heat capacity,
$C$, become positive. Therefore, the heat inflow leads to the 
isothermalization of the core. Secondly, after the maximum expansion
of the core, the region where $C > 0$ decreases, heat starts flowing
from a zone where $C$ is actually now negative. Thirdly, the 
gravothermal instability is
triggered and the core collapses. The system moves from a minimum to 
a maximum entropy state. The time scales of these processes
for monolithic halos depend on \sig. For cosmological halos these
time scales can be modified by extra dynamical heating (e.g., mergers):
the halo periphery can be heated and 
the temperature there can become higher than in the core, 
the heat flows to 
the core and it expands until the overall halo isothermalizes. This 
process can delay the core collapse, but will not reverse the gravothermal 
instability to a runaway core expansion.

\begin{figure*}[!t]
  \includegraphics[width=\textwidth]{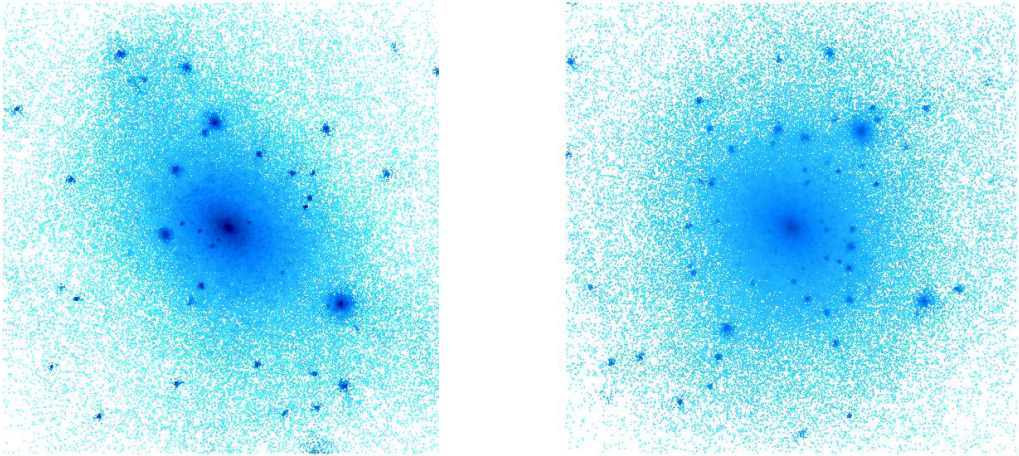}
  \caption{Distribution of dark matter particles inside a box of 0.5 \mpch\
on a side centered on a MW-sized halo with $\sig = 0$ (left, CDM) and with
$\sig = 0.5(1/\vcien)$ (right, SIDM). Particles are color-coded on a gray 
scale according to their local density. The local density at the particle
positions was computed using SMOOTH.
%, a publicly available code developed by the HPCC group in the Department 
of Astronomy of the University of Washington.
}
\end{figure*}

%cosmological halos
We have further investigated Milky Way (MW)- and cluster-sized halos from
cosmological simulations, using several values for the pair 
($\sign,\alpha$). The parameter that defines the evolution under
SI is the number of collisions per particle after a given time,
$\ncol\propto \rho \ \sig $ v$_{\rm rms}\ t$. From all of our runs,
we see that if \ncol\ at the present epoch is $\lesssim 2-5$, then the
halo is either in the core-expansion phase or has just had the 
gravothermal catastrophe triggered. The halo density profiles are
affected (flattened) only in the innermost regions, $r<0.05\rvir$.  
When \sig\ is constant ($\alpha =0$), 
the cores of low velocity halos are on average less influenced by SI 
than the high velocity ones; therefore, \rhosl\ will depend on scale
(but see Dav\'e et al. 2001). When $\sig\propto 1/\vrel$
($\alpha =1$), we expect halos of different sizes to have roughly similar 
core densities (Firmani et al. 2001; Yoshida et al. 2001)
Nevertheless, these simple reasonings apply strictly only 
for monolithic halos. The cosmological mass assembly history may dramatically 
influence the $z=0$ structure of halos with a significant cross section; 
for example,
for values of $\sign = 3\ $cm$^2$gr$^{-1}$ and $\alpha = 0$, 
we find that MW- or cluster-sized halos are or are not well
inside the core collapse phase 
by $z=0$; this depends essentially on whether halos have suffered 
recent major mergers. 
We see clearly that the smooth mass accretion is not important; only the violent
mergers can delay the core collapse phase.

%central densities
We find that only SIDM models with $\sig\propto 1/\vrel$ are able to predict 
halo central densities that do not depend on scale, as observations suggest. 
For  $\sig = 0.5-1.0 (1/\vcien)$cm$^2$gr$^{-1}$, dwarf to 
cluster-sized halos have \rhosl\ values within the range inferred
from observations. In Fig. 1, we present a comparison between a
MW-sized SIDM ($\sig = 0.5(1/\vcien)$cm$^2$gr$^{-1}$) and a CDM halo
(the so-called, MW-sized, fiducial halo in C2002). Slight
differences can be noticed from this figure: (a) the core
of the SIDM halo is less concentrated, the black spot at the center
of the halo is less dark and appears uniformly spread over a larger area, (b)  
small subhalos are slightly more numerous in the SIDM halo and have a different
radial distribution, and (c) the SIDM halo seems to be rounder (a 
quantitative and more detailed comparison can be found in  C2002).

%substructure
In general, we find that the number of subhalos within the SIDM halos 
simulated here is largely suppresed only for ($\sign,\alpha$) = (3.0,0.0).
Subhalos may survive a longer time in SIDM halos than in CDM ones because 
the inner tidal force in the former is weaker than in the latter.
The structure and population of subhalos are determined 
by the interaction between the hot host halo particles and the cooler 
subhalo particles, rather than by internal processes in the subhalos. 
Overall, SIDM subhalos become puffier than their CDM counterparts; they 
are completely evaporated only in the limit of high \sig.

Our simulations also show that SI start flattening the inner density profiles 
of growing CDM halos since early epochs in such a way that the growth of 
super massive black holes by accretion of SIDM  (Hennawi \& Ostriker 2001) 
does not seem to be an efficient process, and therefore 
can not be used as a criterion to constrain SIDM models. 

This work was supported by CONACyT grants 33776-E to V.A. and 
36584-E to P.C.

\end{document}